\documentclass{report}

\usepackage{natbib}

\usepackage{listings}
\usepackage{color}

\usepackage{amsmath}
 
\definecolor{codegreen}{rgb}{0,0.6,0}
\definecolor{codegray}{rgb}{0,0,0.5}
\definecolor{codepurple}{rgb}{0.58,0,0.82}
\definecolor{backcolour}{rgb}{0.99,0.99,0.99}
 
\lstdefinestyle{mystyle}{
    backgroundcolor=\color{backcolour},   
    commentstyle=\color{codegreen},
    keywordstyle=\color{magenta},
    numberstyle=\tiny\color{codegray},
    stringstyle=\color{codepurple},
    basicstyle=\footnotesize,
    breakatwhitespace=false,         
    breaklines=true,                 
    captionpos=b,                    
    keepspaces=true,                 
    numbers=left,                    
    numbersep=5pt,                  
    showspaces=false,                
    showstringspaces=false,
    showtabs=false,                  
    tabsize=2
}
\usepackage[caption = false]{subfig}

\usepackage{float}

\usepackage{algpseudocode}

\usepackage{graphicx}

\usepackage{tabularx}

\makeatletter
\def\BState{\State\hskip-\ALG@thistlm}
\makeatother

\graphicspath{{./Figures/}}

\usepackage{url}

\usepackage{fancyhdr}
\usepackage{fancyhdr}
\usepackage{lastpage}

\usepackage{amsmath}
\usepackage{amssymb}
\usepackage{graphicx}
\usepackage{epstopdf}
\usepackage{inputenc}
\usepackage{geometry}
\title{%
  \Huge OSmOSE report $No$ 1 \\ 
  \LARGE Theory-plus-code documentation of the DEPAM workflow for soundscape description}

\date{}

\begin{document}

\maketitle

%
%
%

\begin{abstract} 
In the Big Data era, the community of PAM faces strong challenges, including the need for more standardized processing tools accross its different applications in oceanography, and for more scalable and high-performance computing systems to process more efficiently the everly growing datasets. In this work we address conjointly both issues by first proposing a detailed theory-plus-code document of a classical analysis workflow to describe the content of PAM data, which hopefully will be reviewed and adopted by a maximum of PAM experts to make it standardized. Second, we transposed this workflow into the Scala language within the Spark/Hadoop frameworks so it can be directly scaled out on several node cluster.
\end{abstract}

\newpage

\paragraph{Authorship}

This document was drafted by Dorian Cazau (Institute Mines Telecom Atlantique) with the assistance of Paul Nguyen (Sorbonne Universit\'es).

\paragraph{Document Review}

Though the views in this document are those of the authors, it was reviewed by a panel of french acousticians before publication. This enabled a degree of consensus to be developed with regard to the contents, although complete unanimity of opinion is inevitably difficult to achieve. Note that the members of the review panel and their employing organisations have no liability for the contents of this document. The Review Panel consisted of the following experts (listed in alphabetical order):

\begin{itemize}
\item Dorian Cazau$^1$ (corresponding author, m: cazaudorian@outlook.fr)
\item Paul Nguyen Hong Duc$^2$
\end{itemize}

belonging to the following institutes (at the time of their contribution)
\begin{enumerate}
\item Institute Mines Telecom Atlantique
\item Sorbonne Universit\'es
\end{enumerate}

\paragraph{Last date of modifications}

\today

\paragraph{Recommended citation} "Theory-plus-code documentation of the DEPAM workflow for soundscape description", OSmOSE report $No$ 1

\paragraph{Future revisions}

Revisions to this guide will be considered at any time. Any suggestions for additional material or modification to existing material are welcome, and should be communicated to Dorian Cazau (cazaudorian@outlook.fr).

\paragraph{Document and code availability}

This document has been made open source under a Creative Commons Attribution-Noncommercial-ShareAlike license (CC BY-NC-SA 4.0) on arxiv (). All associated codes have also been released in open source and access under a GNU General Public License and are available on github (\url{https://github.com/Project-ODE/FeatureEngine}).

\paragraph{Acknowledgements}

All codes have been produced by the developer team of the non-profit association Ocean Data Explorer (w: \url{https://oceandataexplorer.org/}, m: oceandataexplorer@gmail.com). We thank the Pôle de Calcul et de Données pour la Mer from IFREMER for the provision of their infrastructure DATARMOR and associated services. The authors also would like to acknowledge the assistance of the review panel, and the many people who volunteered valuable comments on the draft at the consultation phase.

\tableofcontents

\newpage

\chapter{Introduction}

\section{Context}

Measured noise levels in Passive Acoustic Monitoring (PAM) are sometimes difficult to compare because different measurement methodologies or acoustic metrics are used, and results can take on different meanings for each different application, leading to a risk of misunderstandings between scientists from different PAM disciplines. For reasons of comparability, and since it is cumbersome to define each term every time it is used, some common definitions are needed for acoustic metrics. 

In the hope of boosting standardization and interoperability, numerous efforts have already been made to outline some best practices regarding PAM both as an ocean observing measure and as a STIC discipline. \cite{Robinson2014} provided a full technical report of best practices, reviewed by a comitee of experts. \cite{Merchant2015} provided a comprehensive overview of PAM methods to characterize acoustic habitats, and released an open-source toolbox both in R and Matlab with a theoretical document.


\section{Contributions}

In the same vein, our work addresses the need for a common approach, and the desire to promote best practices for processing the data, and for reporting the measurements using appropriate metrics. 

We release a new open source end-to-end analytical workflow for description and interpretation of underwater soundscapes, along with the present document. We outline the following contributions 

\begin{itemize}
\item this workflow has been implemented in \textbf{three different computer languages}: Matlab, Python and Scala. These three implementations perfectly match in regards to the unitary tests done on core functions, with rms error below 10$^{-16}$, and to the data processing operations and end-user functionalities and results. Note also that in these implementations we try at best to fit with "the best practices in programming" from the DCLDE community in Passive Acoustic Monitoring, for the Matlab implementation, and with the web community and data scientists, for the Scala implementation. These different versions of the workflow have been released on github under a GNU licence;

\item in this document, we aligned the lines of codes with their corresponding theoretical signal processing definitions, so as to \textbf{fill at best the gap between theory and code};

\item the Scala implementation of the workflow allows for a \textbf{direct and transparent scaling out of data processing} over a CPU cluster using the Hadoop/Spark frameworks, allowing for significant computational gain.

\end{itemize}

As stated in the preamble, this workflow has been collaboratively elaborated, co-developed and reviewed by a research team gathering more than 2 PAM experts over 2 different institutes. Thus, it should provide a reliable value of standardization. Also, during all our work, we built at best on similar works in order to avoid replicating previous efforts. In table \ref{SourceCodeReviewed}, we list the different source codes on which we have relied to implement our workflow. In reference to these sources, we systematically highlighted agreements and disagreements with their implementations (and theoretical explanations when present) in the paragraphs named ``Discussion", discussed them in regards to each of these different sources and thus justified the choices made for our own implementation.

Eventually, note that reported codes in this document are not representative of their real implementation structure (e.g. in terms of functions), but we rather focus on reporting the essential code lines that implement litteraly each equation and theoretical points.

\begin{table}[htbp]
\centering
\resizebox{\columnwidth}{!}{
\begin{tabular}{|c|c|c|c|}
\hline
Code source    & Language                & Main functions used  & References \\ \hline

Package scipy v-1.0.0  & Python                     & stft.py / spectrogram.py / welch.py  &   \url{https://www.scipy.org/} \\ \hline

Matlab 2014a & Matlab    & spectrogram.m / pwelch.m  & MathWorks   \\ \hline

pamGuide & R / Matlab &  PAMGuide.R  & \cite{Merchant2015}  \\ \hline
\end{tabular}
}
\caption{Details of codes reviewed.}
\label{SourceCodeReviewed}

\end{table}

\vspace{0.3cm}

\section{Overview}

\begin{figure}[htbp]
\centering
\includegraphics[width=0.65\columnwidth]{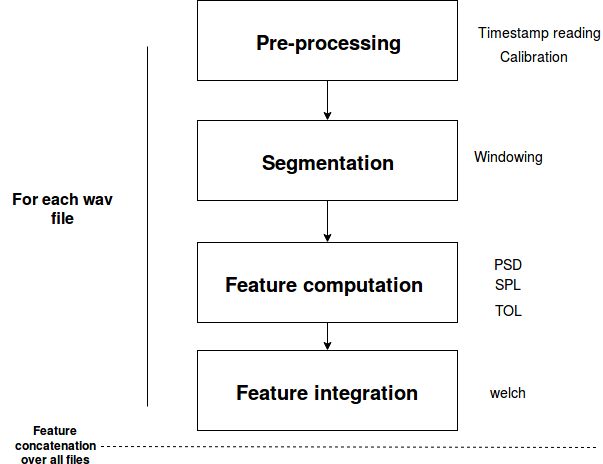}
\caption{Diagram block.}\label{DiagBlock}
\end{figure}

As shown in figure \ref{DiagBlock}, our workflow is composed of the following blocks
\begin{itemize}
\item pre-processing (Sec. \ref{preProcessing});
\item segmentation (Sec. \ref{Segmentation});
\item feature computation and integration (Sec. \ref{FeatureComputationAndIntegration});
\end{itemize}

Note that we have two different time scales for data analysis: 
\begin{itemize}
\item first scale (see Section \ref{Segmentation}): for feature computation in short-term analysis windows of length ``windowSize``;
\item second scale (see Section \ref{FeatureComputationAndIntegration}): for feature integration in longer time segments, applied when $segmentSize > windowSize$.
\end{itemize}

Note that when $segmentSize <= windowSize$, these time scales are similar and only one segmentation is performed.

\vspace{0.3cm}

The implemented acoustic metrics are (selected among the list in \cite[Sec. 2.1.2]{Robinson2014})
\begin{itemize}
\item \textbf{PSD} Power Spectral Density;
\item \textbf{TOL} Third-Octave Levels;
\item \textbf{SPL} Sound Pressure Level
\end{itemize}

%
%
%
%
%
%

\chapter{Pre-processing}\label{preProcessing}

\section{Timestamp reading}

\subsection{Theory}
 
The CSV file must contain (at least) the following columns: 
\begin{itemize}
\item filename: "Example0\_16\_3587\_1500.0\_1.wav"	
\item start\_date: "2010-01-01T00:00:00Z"
\end{itemize}
 
The workflow first imports the list of filenames and only process corresponding audio files. Thus, an audio file not referred into the csv file will not be processed. Note that this metadata organization corresponds to the raw format of several manufacturers of recorders such as AURAL.

\subsection{Matlab code}

\paragraph{Correspondences with theory}

Reading the list of filenames from csv is performed at line 3. The structure of audio file metadata is enforced at lines 5-9. No more detailed explanations needed.

\begin{lstlisting}[language=Matlab]
fid = fopen('../../test/resources/metadata/Example_metadata.csv');
metadataHeader = textscan(fid,'%q %q', 1, 'delimiter', ',');
metadata = textscan(fid,'%q %q','delimiter',',');
fclose(fid);
wavFiles = struct(...
    'name', string(metadata{1}),...
    'fs', [1500, 1500],...
    'date', string(metadata{2})...
);
\end{lstlisting}

\paragraph{Discussion}

No sources, custom code.

\subsection{Python code}

\paragraph{Correspondences with theory}

Reading the list of filenames from csv is performed at line 8. The structure of audio file metadata is enforced at lines 1-7. No more detailed explanations needed.

\begin{lstlisting}[language=Python]
FILES_TO_PROCESS = [{
    "name": file_metadata[0],
    "timestamp": parse(file_metadata[1]),
    "sample_rate": 1500.0,
    "wav_bits": 16,
    "n_samples": 3587,
    "n_channels": 1
} for file_metadata in pd.read_csv(METADATA_FILE_PATH).values]
\end{lstlisting}

\paragraph{Discussion}

No sources, custom code.

\section{Audio reading and calibration}\label{Calibration-Code}

\subsection{Theory}

Initially, $xin$ is a digital (bit-scaled) audio signal recorded by the hydrophone, such that the amplitude range is -2$^{N_{bit}-1}$ to 2$^{N_{bit}-1}$-1. A first calibration operation is to convert this signal into a time-domain acoustic pressure signal (also called pressure waveform, in Pa, as defined by the International System of Units) as follows:

\begin{equation}\label{Calib1}
xin = \frac{xin}{10^{\frac{S}{20}}}  \quad [Pa]
\end{equation}
 
where $S$ is the calibration correction factor corresponding to the hydrophone sensitivity (typically in dB ref 1 V/ $\mu$ Pa, with negative values for underwater measurements). Note that it is possible to correct for the variation in the sensitivity with frequency if the hydrophone is calibrated over the full frequency range of interest [IEC 60565 2006]. When this factor is frequency dependent, it must be applied within spectral features (see eq 10, 16 and 17 in \cite[Appendix 1]{Merchant2015}).

\subsection{Matlab code}

\paragraph{Correspondences with theory} Eq. \ref{Calib1} is performed in line 2.

\begin{lstlisting}[language=Matlab]
rawSignal=audioread(strcat(wavFileLocation, wavFileName),'double');
calibratedSignal = rawSignal * (10 ^ (calibrationFactor / 20));
\end{lstlisting}

\paragraph{Discussion}

Used in the function PG\_Waveform.m from PAMGuide \citep[eq. 21]{Merchant2015}.

\subsection{Python code}

\paragraph{Correspondences with theory} Eq. \ref{Calib1} is performed in line 2.

\begin{lstlisting}[language=Python]
sound, sample_rate = self.sound_handler.read()
calibrated_sound = sound / 10 ** (self.calibration_factor / 20)  
\end{lstlisting}

\paragraph{Discussion}

No sources, custom code.

\chapter{Segmentation}\label{Segmentation}

\section{Case where $segmentSize > windowSize$}

\subsection{Theory}

We call segmentation the division of the time-domain signal, $x$, into $segmentSize$-long segments. The s$^{th}$ segment is given by

\begin{equation}\label{Seg1}
segment^s[n] = xin[n + mN]
\end{equation}

where $N$ is the number of samples in each window, 0 $\leq$ n $\leq$ N-1 \citep{Marple1987} and  0 $\leq$ s $\leq$ S. For each audio file, a certain number of segments S is obtained, and the last truncated one is removed.

We then perform a short-term division of each segment $segment$ into $windowSize$-long windows, which may be overlapping in time. The m$^{th}$ window is given by

\begin{equation}\label{Seg2}
xin^m[n] = segment[n + (1 - r)mN]
\end{equation}

where $N$ is the number of samples in each window, 0 $\leq$ n $\leq$ N-1 \citep{Marple1987}, $r$ is the window overlap and $M$ is the number of windows in a segment. The last truncated short-term window is removed. A window function is then applied to each data chunk. Denoting the $m^{th}$ windowed data chunk $xin^{(m)}_{win}[n]$

\begin{equation}\label{Wind1}
xin_{win}^{(m)}[n] = \frac{w[n]}{\alpha} xin^{(m)}[n]
\end{equation}

where $w$ is the window function over the range 0 $\leq$ n $\leq$ $N$-1, and $\alpha$ is the scaling factor, which corrects for the reduction in amplitude introduced by the window function \citep{Cerna2000}.

\paragraph{Discussion}

This section has been drawn from \cite[Supplementary Material]{Merchant2015}. However, we introduce two successive levels of segmentation, integration-level and short-term window-level, where the second is imbricated into the first one. We follow here the order of segmentations as they appear in numerical implementations, making explicit the truncation problem when $windowSize$ is not an integer multiple of $segmentSize$, which is not transparent in the paragraph of \cite[Supplementary Material, sectin 6.4]{Merchant2015}.

\section[$segmentSize <= windowSize$]{Case where $segmentSize <= windowSize$}\label{Calibration-Code}

\subsection{Theory}

In this case, only the short-term segmentation into analysis windows is performed (ie eq. \ref{Seg2} and \ref{Wind1}), only now the segment is seen as the full audio file, so that $M$ (in eq. \ref{Seg2}) is the number of windows into the complete audio file. Likewise, the last truncated short-term window is removed.

\paragraph{Discussion}

This section has been drawn from \cite[Supplementary Material]{Merchant2015} without any modifications.

\subsection{Matlab code}

\paragraph{Correspondences with theory} 

After variable initialization (lines 1-3), eq. \ref{Seg1} is done at line 8 and eq. \ref{Wind1} at line 13. The scaling factor $\alpha$ is included in the variable windowFunction.

\begin{lstlisting}[language=Matlab]
segmentSize = fix(segmentDuration * fs);
nSegments = fix(wavInfo.TotalSamples / segmentSize);
windowFunction = hamming(windowSize, 'periodic'); 

% going backwards to have the right struct size allocation of results
for iSegment = nSegments-1 : -1 : 0
    
    signal = calibratedSignal(1 + iSegment*segmentSize : (iSegment+1) * segment Size);
                            
    nPredictedWindows = fix((length(signal) - windowOverlap) / (windowSize - 		windowOverlap));

    % grid whose rows are each (overlapped) segment for analysis
    segmentedSignalWithPartial = buffer(signal, windowSize, windowOverlap, 'nodelay');

    segmentedSignalWithPartialShape = size(segmentedSignalWithPartial);

    % remove final segment if not full
    if segmentedSignalWithPartialShape(2) ~= nPredictedWindows
        segmentedSignal = segmentedSignalWithPartial(:, 1:nPredictedWindows);
    else
        segmentedSignal = segmentedSignalWithPartial;
    end
    
	% multiply segments by window function
	windowedSignal = bsxfun(@times, segmentedSignal, windowFunction);
    
        
    %% FEATURE COMPUTATION 
    
\end{lstlisting}

\paragraph{Discussion}

Drawn from the function pwelch.m in Matlab 2014a.

\subsection{Python code}

\paragraph{Correspondences with theory} 

After variable initialization (line 1), eq. \ref{Seg1} is done at line 2-4 and eq. \ref{Wind1} at lines 5-6. The scaling factor $\alpha$ is included in the function win.

\begin{lstlisting}[language=Python]
nSegments = sound.shape[0] // self.segmentSize
segmentedSound = numpy.split(sound[:self.segmentSize * nSegments], nSegments)
for iSegment in range(nSegments):

	signal=segmentedSound[iSegment]            
	shape = (nWindows, windowSize)
    strides = (nWindows * signal.strides[0], signal.strides[0])
    windows = np.lib.stride_tricks.as_strided(signal, shape=shape, strides=strides)
    windowedSignal = windows * windowFunction
    
    %% FEATURE COMPUTATION 
        
\end{lstlisting}

\paragraph{Discussion}

Adapted from the function spectrogram in scipy, modifications only done to make this code suitable for our variable names.

\chapter{Feature Computation}\label{FeatureComputationAndIntegration}

\section{PSD (Power Spectral density)}

\subsection{Theory}

The Discrete Fourier Transform (DFT) of the $m^{th}$ segment $X^{(m)}(f)$ is given by

\begin{equation}\label{DefDFT}
X^{(m)}(f) = \sum_{n=0}^{N-1} xin_{win}^{(m)}[n] e^{\frac{-i 2\pi f n}{N}}
\end{equation}

The power spectrum is computed from the DFT, and corresponds to the square of the amplitude spectrum (DFT divided by N), which for the $m^{th}$ segment is given by

\begin{equation}\label{Pdef}
P^{(m)}(f) = |\frac{X^{(m)}(f)}{N}|^2
\end{equation}

where $P^{(m)}(f)$ stands for the power spectrum. For real sampled signals, the power spectrum is symmetrical around the Nyquist frequency, $Fs/2$, which is the highest frequency which can be measured for a given $Fs$. The frequencies above $Fs/2$ can therefore be discarded and the power in the remaining frequency bins are doubled, yielding the single-sided power spectrum

\begin{equation}\label{Pdouble}
P^{(m)}(f') = 2 . P^{(m)}(f')
\end{equation}

where $0 < f' < fs/2$. This correction ensures that the amount of energy in the power spectrum is equivalent to the amount of energy (in this case the sum of the squared pressure) in the time series. This method of scaling, known as Parseval's theorem, ensures that measurements in the frequency and time domain are comparable. The power spectral density $PSD$ (also called mean-square sound-pressure spectral density) is defined by:

\begin{equation}\label{NormaPSD}
PSD(f',m) = \frac{P^{(m)}(f')}{B \Delta f} \quad \textrm{[$\mu$Pa$^2$ /Hz]}
\end{equation}

where $\Delta f = fs/2N$ is the width of the frequency bins, and $B$ is the noise power bandwidth of the window function, which corrects for the energy added through spectral leakage:

\begin{equation}
B = \frac{1}{N} \sum_{n=0}^{N-1} (\frac{w[n]}{\alpha})^2
\end{equation}

Note that a spectral density is any quantity expressed as a  contribution per unit of bandwidth. A spectral density level is ten times the logarithm to the base 10 of the ratio of the spectral density of a quantity per unit bandwidth, to a reference value. Here the power spectral density level would be expressed in units of dB re 1 $\mu$Pa$^2$ /Hz.

\paragraph{Discussion}

This section has been integrally drawn from \cite[Supplementary Material]{Merchant2015} without any modifications.

\subsection{Matlab code}

\paragraph{Correspondences with theory} Eq. \ref{DefDFT} is performed at lines 6-7. Eq. \ref{Pdef} is performed at lines 8. Eq. \ref{Pdouble} is performed at lines 9.

\begin{lstlisting}[language=Matlab]
if (mod(nfft, 2) == 0)
	spectrumSize = nfft/2 + 1;
else
    spectrumSize = nfft/2;
end
twoSidedSpectrum = fft(windowedSignal, nfft);
oneSidedSpectrum = twoSidedSpectrum(1 : spectrumSize, :);
powerSpectrum = abs(oneSidedSpectrum) .^ 2;
powerSpectrum(2 : spectrumSize-1, :) = powerSpectrum(2 : spectrumSize-1, :) .* 2;
psdNormFactor = 1.0 / (fs * sum(windowFunction .^ 2));
powerSpectralDensity = powerSpectrum * psdNormFactor;
welch = mean(powerSpectralDensity, 2);
\end{lstlisting}

\paragraph{Discussion}

Drawn from the function pwelch.m in Matlab 2014a.

\subsection{Python code}

\paragraph{Correspondences with theory} Eq. \ref{DefDFT} is performed at lines 1-3. Eq. \ref{Pdef} is performed at lines 4-7. Eq. \ref{Pdouble} is performed at lines 8-13. Eq. \ref{NormaPSD} is performed at lines 14-16.

\begin{lstlisting}[language=Python]
rawFFT = np.fft.rfft(windowedSignal, nfft)
vFFT = rawFFT * np.sqrt(1.0 / windowFunction.sum() ** 2)
periodograms = np.abs(rawFFT) ** 2
vPSD = periodograms / (fs * (windowFunction ** 2).sum())
vWelch = np.mean(vPSD, axis=0)
\end{lstlisting}

\paragraph{Discussion}

Adapted from the function spectrogram in scipy, with modifications only done to make this code suitable for our variable names.

\section{TOL (Third-Octave Levels)}

\subsection{Theory}

Center frequencies can be computed in base-two and base-ten. In our computations, only base-ten exact center frequencies were used. It has to be noted that the nominal frequency is not the exact value of the corresponding center frequency. Readers are referred to \cite{wiki-TOL} and ISO standards to have the first center frequencies of the TOLs. Center frequencies of the TOLs can be calculated as follow:
\begin{equation}\label{TO}
toCenter = 10^{0.1*i}
\end{equation}
with $i$ the number of the TOL. In order to determine the bandedge frequencies of each TOL, ANSI and ISO standards give the following equations:
\begin{equation}\label{bandedgeFreqs}
\begin{split}
lowerBoundFrequency & = toCenter \div tocScalingFactor \\
upperBoundFrequency & = toCenter \times tocScalingFactor
\end{split}
\end{equation}
with $toCenter$ the center frequency of the TOL and $tocScalingFactor = 10^{0.05}$. From \cite[Appendix 1]{Merchant2015} and \cite{Richardson1995}, a TOL is defined as the sum of the sound powers within all 1-Hz bands included in the third octave band (third octave band). Mathematically, according to \cite[Supplementary Material]{Merchant2015}, it can be expressed as:

\begin{equation}\label{TOL}
TOL(toCenter) = 10 log_{10}(\frac{1}{p_{ref}^2} \sum_{f=lowerBoundFrequency}^{f=upperBoundFrequency} \frac{P(f)}{B}) - S(toCenter)
\end{equation}

For computational efficiency, TOLs are computed by summing the frequency bins of the power spectrum that are included in a TOL. In \cite{ISO266-1975} and \cite{ANSIS1.11-2004} standards, filters with specific characteristics should be designed to compute TOLs with the time-domain signal. For what concerns TOL units, \cite{Richardson1995} and \cite[Supplementary Material]{Merchant2015} disagree about units. For \cite{Richardson1995}, correct units are dB re 1 $\mu$Pa whereas for \cite[Supplementary Material]{Merchant2015}, TOL units are dB re 1 $\mu$Pa or dB re 1 $\mu$Pa$^2$ or dB. Note that for accurate representation of third-octave band levels at low frequencies, a long snapshot time is required (sufficient accuracy at 10 Hz requires a snapshot time of at least 30 seconds).

\subsection{Matlab code}

\paragraph{Correspondences with theory} 

All these conditions are to be met in order to follow the ISO and ANSI standards. TOL are computed for a second and Nyquist frequency cannot be exceeded. Moreover, we have chosen to start our TOL computations with the TOB at 1Hz. However, we are aware that the TOBs under 25 Hz lead to inaccurate computations (Mennitt and Fristrup, 2012). This can be easily modified in that condition $if (lowFreqTOL < 1.0)$.

\begin{lstlisting}[language=Matlab]
    if (length(signal) < sampleRate)
        MException('tol:input', ['Signal incompatible with TOL computation, '...
        'it should be longer than a second.'])
    end

    if (length(windowFunction) ~= sampleRate)
        MException('tol:input', ['Incorrect windowFunction for TOL, '...
        'it should be of size sampleRate.'])
    end

    if (lowFreqTOL < 1.0)
        MException('tol:input', ['Incorrect lowFreq for TOL, '...
        'it should be higher than 1.0.'])
    end

    if (highFreqTOL > sampleRate/2)
        MException('tol:input', ['Incorrect highFreq for TOL, '...
        'it should be lower than sampleRate/2.'])
    end

    if (lowFreqTOL > highFreqTOL)
        MException('tol:input', ['Incorrect lowFreq,highFreq for TOL, '...
        'lowFreq is higher than highFreq.'])
    end
    
\end{lstlisting}
After the normalized power spectrum computation, the TOL calculation is done.
Eq. \ref{TO} and Eq. \ref{bandedgeFreqs} are done in the following code:
\begin{lstlisting}[language=Matlab]
    tobCenters = 10 .^ ((0:59) / 10);

    tobBounds = zeros(2, 60);
    tobBounds(1, :) = tobCenters * 10 ^ -0.05;
    tobBounds(2, :) = tobCenters * 10 ^ 0.05;
\end{lstlisting}

We chose to set the TOB centers in order to ba as close as possible to the Scala workflow to have a consistent benchmark. However, in PAMGuide, the TOB centers are set according to the frequency range set by the user. The 59th TOB center corresponds to about 794328 Hz which is much more greater than standard sampling rate of hydrophones. It has to be noted that this value can also be easily modified.
Eq. \ref{TOL} is done in the following code:
\begin{lstlisting}[language=Matlab]
    % Find indices of the TOB 
    inRangeIndices = find((tobBounds(2, :) < sampleRate / 2)...
        & (lowFreqTOL <= tobBounds(2, :))...
        & (tobBounds(1, :) < highFreqTOL));
    % Convert indices to match those in the spectrum
    tobBoundsInPsdIndex = zeros(2, length(inRangeIndices));
    tobBoundsInPsdIndex(1, :) = fix(tobBounds(1, inRangeIndices(1):inRangeIndices(end)) * (nfft / sampleRate));
    tobBoundsInPsdIndex(2, :) = fix(tobBounds(2, inRangeIndices(1):inRangeIndices(end)) * (nfft / sampleRate));

    tol = zeros(1, length(inRangeIndices));
    % Compute TOL
    for i = 1 : length(inRangeIndices)
        tol(i) = sum(sum(...
            normalizedPowerSpectrum(1+tobBoundsInPsdIndex(1, i) : tobBoundsInPsdIndex(2, i), :)...
        , 1));
    end

    tol = 10 * log10(tol);
\end{lstlisting}

Eq. \ref{TO} is done with the or loop in the following code:

\begin{lstlisting}[language=Matlab]
% Calculate centre frequencies (corresponds to Eq. 4.6 in the User doc and 13 in PAMGuide tutorial)        
for i = 2:nband                 %calculate 1/3 octave centre 
    fc(i) = fc(i-1)*10^0.1;     % frequencies to (at least) precision 
end                             % of ANSI standard
\end{lstlisting}

Eq. \ref{bandedgeFreqs} is done at lines 2 and 3: 

\begin{lstlisting}[language=Matlab]
% Calculate boundary frequencies of each band (EQUATIONS 14-15 in PAMGuide tutorial and 4.7 in User doc)    
fb = fc*10^-0.05;               %lower bounds of 1/3 octave bands
fb(nfc+1) = fc(nfc)*10^0.05;    %upper bound of highest band (upper
                                %   bounds of previous bands are lower
                                %   bounds of next band up in freq.)
if max(fb) > hcut               %if highest 1/3 octave band extends 
    nfc = nfc-1;                %   above highest frequency in DFT, 
end
\end{lstlisting}

Eq. \ref{TOL} is done in the following code:
\begin{lstlisting}[language=Matlab]
% Calculate 1/3-octave band levels (corresponds to EQUATION 16 in PAMGuide tutorial and 4.8 in the User doc)
    P13 = zeros(M,nfc);             %initialise TOL array
        
    for i = 1:nfc                   %loop through centre frequencies
        fli = find(f >= fb(i),1,'first');   %index of lower bound of band
        fui = find(f < fb(i+1),1,'last');   %index of upper bound of band
        for q = 1:M                 %loop through DFTs of data segments
            fcl = sum(Pss(q,fli:fui));%integrate over mth band frequencies
            P13(q,i) = fcl ;         %store TOL of each data segment
        end
    end
    if ~isempty(P13(1,10*log10(P13(1,:)/(pref^2)) <= -10^6))
        lowcut = find(10*log10(P13(1,:)/(pref^2)) <= -10^6,1,'last') + 1;
                                    %index lowest band before empty bands
                                    % at low frequencies
        P13 = P13(:,lowcut:nfc);        %remove empty low-frequency bands
    end
	a = 10*log10((1/B)*P13/(pref^2))-S; %TOLs
clear P13
clear Pss

%% Construct output array
A = 10*log10(mean(10.^(double(a)./10))); % Mean aggregation depending on the length of integration windows
\end{lstlisting}

\paragraph{Discussion}

Dranw from PAMGuide \citep{Merchant2015}.

\subsection{Python code}

\paragraph{Correspondences with theory} 
All these conditions are to be met in order to follow the ISO and ANSI standards as in Matlab codes.

\begin{lstlisting}[language=Python]
# We're using some accronymes here:
#   toc: third octave center
#   tob: third octave band
        if nfft is not int(sample_rate):
            Exception(
                "Incorrect fft-computation window size ({})".format(nfft)
                + "for TOL (should be higher than {})".format(sample_rate)
            )

        self.lower_limit = 1.0
        self.upper_limit = max(sample_rate / 2.0,
                               high_freq if high_freq is not None else 0.0)

        if low_freq is None:
            self.low_freq = self.lower_limit
        elif low_freq < self.lower_limit:
            Exception(
                "Incorrect low_freq ({}) for TOL".format(low_freq)
                + "(lower than lower_limit{})".format(self.lower_limit)
            )
        elif high_freq is not None and low_freq > high_freq:
            Exception(
                "Incorrect low_freq ({}) for TOL".format(low_freq)
                + "(higher than high_freq {}".format(high_freq)
            )
        elif high_freq is None and low_freq > high_freq:
            Exception(
                "Incorrect low_freq ({}) for TOL".format(low_freq)
                + "(higher than upper_limit {}".format(self.upper_limit)
            )
        else:
            self.low_freq = low_freq

        if high_freq is None:
            self.high_freq = self.upper_limit
        elif high_freq > self.upper_limit:
            Exception(
                "Incorrect high_freq ({}) for TOL".format(high_freq)
                + "(higher than upper_limit {})".format(self.upper_limit))
        elif low_freq is not None and high_freq < low_freq:
            Exception(
                "Incorrect high_freq ({}) for TOL".format(low_freq)
                + "(lower than low_freq {})".format(high_freq)
            )
        elif low_freq is None and high_freq < self.lower_limit:
            Exception(
                "Incorrect high_freq ({}) for TOL".format(high_freq)
                + "(lower than lower_limit {})".format(self.lower_limit)
            )
        else:
            self.high_freq = high_freq

        # when wrong low_freq, high_freq are given,
        # computation falls back to default values
        if not self.lower_limit <= self.low_freq\
                < self.high_freq <= self.upper_limit:

            Exception(
                "Unexpected exception occurred - "
                + "wrong parameters were given to TOL"
            )

        self.sample_rate = sample_rate
        self.nfft = nfft

        self.tob_indices = self._compute_tob_indices()
        self.tob_size = len(self.tob_indices)
\end{lstlisting}
Eq. \ref{TO} and Eq. \ref{bandedgeFreqs} are done in the following code:
\begin{lstlisting}[language=Python]
    def _compute_tob_indices(self):
        max_third_octave_index = floor(10 * log10(self.upper_limit))

        tob_center_freqs = np.power(
            10, np.arange(0, max_third_octave_index + 1) / 10
        )

        all_tob = np.array([
            _tob_bounds_from_toc(toc_freq) for toc_freq in tob_center_freqs
        ])

        tob_bounds = np.array([
            tob for tob in all_tob
            if self.low_freq <= tob[1] < self.upper_limit
            and tob[0] < self.high_freq
        ])

        return np.array([self._bound_to_index(bound) for bound in tob_bounds])

    def _bound_to_index(self, bound):
        return np.array([floor(bound[0] * self.nfft / self.sample_rate),
                         floor(bound[1] * self.nfft / self.sample_rate)],
                        dtype=int)
                        
    def _tob_bounds_from_toc(center_freq):
    return center_freq * np.power(10, np.array([-0.05, 0.05]))
\end{lstlisting}
Eq. \ref{TOL} is done in the following code:
\begin{lstlisting}[language=Python]
    def compute(self, psd):
        third_octave_power_bands = np.array([
            np.sum(psd[indices[0]:indices[1]]) for indices in self.tob_indices
        ])
        return 10 * np.log10(third_octave_power_bands)
\end{lstlisting}

\paragraph{Discussion}

To our knowledge, this is is the first Python version of a TOL computation under the ISO and ANSI standards.

\section{Sound Pressure Levels}

\subsection{Theory}

Sound Pressure Level (SPL), actually the broadband SPL here, is computed as the sum of PSD over all frequency bins, that is

\begin{equation}\label{SPLfilt}
SPL = 10 log_{10}(\frac{1}{B p_{ref}^2} \sum_{f=1}^{nfft} P(f)) 
\end{equation}

with $P$ the single-sided power spectrum (eq. \ref{Pdouble}), $p_{ref} = 1 \mu$ Pa, and $B$ the noise power bandwidth of the window function ($B$=1.36 for a Hamming window).

\paragraph{Discussion}

This section has been integrally drawn from \cite[Supplementary Material, eq. 17]{Merchant2015} without any modifications.

\subsection{Matlab code}

\paragraph{Correspondences with theory} Eq. \ref{SPLfilt} is performed at lines 1

\begin{lstlisting}[language=Matlab]
SPL = 10*log10(mean(vPSD_int))
\end{lstlisting}

\paragraph{Discussion}

No source code has been found for this implementation.

\subsection{Python code}

\paragraph{Correspondences with theory} Eq. \ref{SPLfilt} is performed at lines 1

\begin{lstlisting}[language=Python]
spl = numpy.array([10 * numpy.log10(numpy.sum(welch))])
\end{lstlisting}

\paragraph{Discussion}

No source code has been found for this implementation.

\chapter{Feature integration}\label{FeatureIntegration}

Feature integration is performed in the case where $segmentSize > windowSize$. Note that the timestamp associated with each segment corresponds to the absolute time of the first audio sample in each segment.

\section{Welch}

\subsection{Theory}

When averaging noise, it is necessary first to square the data (since sound pressure has both positive and negative excursions, the unsquared data will tend to average to zero). Therefore, the noise values are most often stated as mean square values, or in terms of root mean square (RMS) values. The Welch method (Welch, 1967) simply consists in time-averaging the M PSD from each segment. The resulting representation consists of the mean of M full-resolution segments averaged in linear space. 

Note that many other averaging operators (eg median) can be used as detailed in \cite[Sec. 5.4.4]{Robinson2014}.

\subsection{Matlab code}

\paragraph{Correspondences with theory} The averaging of PSD is done at the end of each loop (line 4, algorithm 3.2.2).

\begin{lstlisting}[language=Matlab]
vWelch = mean(vPSD, 2)
\end{lstlisting}

\paragraph{Discussion}

No source code has been found for this implementation. Note that Matlab uses a ``datawrap" technique that time-averages analysis window and computes only one single FFT in each segment.

\subsection{Python code}

\paragraph{Correspondences with theory} The averaging of PSD is done at the end of each loop (line 4, algorithm 3.2.3).

\begin{lstlisting}[language=Python]       
vWelch = np.mean(vPSD, axis=0)
\end{lstlisting}

\paragraph{Discussion}

This code has been drawn from the welch function of the scipy package.

\bibliography{/home/cazaudo/Documents/Biblio/References_Biblio}
\bibliographystyle{Cazau}

\end{document}